\documentclass[sigconf]{acmart}
\AtBeginDocument{%
  }

\setcopyright{acmlicensed}
\copyrightyear{2018}
\acmYear{2018}
\acmDOI{XXXXXXX.XXXXXXX}
\acmConference[ ]{ }{ }
\usepackage{color,soul}
\usepackage[framemethod=tikz]{mdframed}
\usepackage{tcolorbox}
\usepackage{diagbox}
\usepackage{tabularray}
\usepackage{multirow}
\usepackage{tikz}
\usepackage{ragged2e}
\usepackage{xcolor}
\usepackage{balance}
\usepackage{enumitem}
\usepackage{graphicx}

\renewcommand\footnotetextcopyrightpermission[1]{} 
\begin{document}

\title{Human or LLM? A Comparative Study on Accessible Code Generation Capability}


\author{Hyunjae Suh}
\affiliation{%
  \institution{University of California, Irvine}
  \city{Irvine}
  \state{CA}
  \country{USA}}
\email{hyunjas@uci.edu}

\author{Mahan Tafreshipour}
\affiliation{%
  \institution{University of California, Irvine}
  \city{Irvine}
  \state{CA}
  \country{USA}}
\email{mtafresh@uci.edu}

\author{Sam Malek}
\affiliation{%
  \institution{University of California, Irvine}
  \city{Irvine}
  \state{CA}
  \country{USA}}
\email{malek@uci.edu}

\author{Iftekhar Ahmed}
\affiliation{%
  \institution{University of California, Irvine}
  \city{Irvine}
  \state{CA}
  \country{USA}}
\email{iftekha@uci.edu}

\renewcommand{\shortauthors}{Suh et al.}

\begin{abstract}
Web accessibility is essential for inclusive digital experiences, yet the accessibility of LLM-generated code remains underexplored. This paper presents an empirical study comparing the accessibility of web code generated by GPT-4o and Qwen2.5-Coder-32B-Instruct-AWQ against human-written code. Results show that LLMs often produce more accessible code, especially for basic features like color contrast and alternative text, but struggle with complex issues such as ARIA attributes. We also assess advanced prompting strategies (Zero-Shot, Few-Shot, Self-Criticism), finding they offer some gains but are limited. To address these gaps, we introduce \textit{FeedA11y}, a feedback-driven ReAct-based approach that significantly outperforms other methods in improving accessibility. Our work highlights the promise of LLMs for accessible code generation and emphasizes the need for feedback-based techniques to address persistent challenges. We provide the source code and datasets that were used in our experiments in the companion website \cite{replicationpackage}.
\end{abstract}

\maketitle

\section{Introduction}
\label{sec:intro}

Web accessibility involves designing and developing websites to ensure they are usable by individuals with diverse (dis)abilities~\cite{w3IntroductionAccessibility}. With approximately 15\% of the global population living with some form of disability~\cite{whoWorldReport}, ensuring web accessibility is crucial for providing equal access to all users. Beyond being an ethical responsibility, it is also a legal obligation. The Americans with Disabilities Act (ADA), enforced by the U.S. Department of Justice, mandates that public-facing businesses and government entities provide accessible web content~\cite{adaGuidanceAccessibility}. As such, web accessibility is no longer optional but a legal and ethical requirement~\cite{peters2010web, lazar2019web}.

Prior studies have investigated various aspects influencing web accessibility, including software frameworks~\cite{angkananon2015technology, heron2013access}, adaptive user interfaces~\cite{raufi2015methods, engel2019svgplott} and design considerations ~\cite{ferati2016web, johari2013web, rothberg2019designing, idrobo2017accessibility} for people with impairments. Additionally, the World Wide Web Consortium (W3C) offers the Web Content Accessibility Guidelines (WCAG)~\cite{wcag}, which provide recommendations for improving web accessibility. Despite this, generating more accessible web content remains a growing focus, with significant ongoing research and efforts dedicated to advancing web accessibility~\cite{mack2021we}.

The recent rise of Large Language Models (LLMs) has profoundly impacted the software development process. Models such as GPT-4o~\cite{gpt4o}, DeepSeek-R1~\cite{deepseekai2025deepseekr1incentivizingreasoningcapability}, Llama3~\cite{llama3}, Qwen2.5-Coder~\cite{hui2024qwen25codertechnicalreport}, and Copilot~\cite{copilot} have demonstrated exceptional performance in tasks including code generation~\cite{ren2023misuse, liu2023codegen4libs, li2022competition, yu2024codereval} and code repair~\cite{zhang2023critical, haque2023potential, sobania2023analysis}. Tools like GitHub Copilot~\cite{copilot} leverage LLMs to assist developers by suggesting code snippets, completing functions, and even generating entire blocks of code from minimal input. Consequently, LLMs are transforming the creation and maintenance of software~\cite{fan2023large}.

As LLM use is becoming evermore prevalent in software development, assessing the LLM-generated code quality has become crucial. Previous research has examined various quality aspects, including the functionality of LLM-generated code, to ensure it performs as intended~\cite{liu2024your, jiang2024surveylargelanguagemodels, li2022competition}. Additionally, studies on code efficiency~\cite{qiu2024efficient, du2024mercury} and security~\cite{siddiq2023generate} have evaluated the overall performance and safety of the code produced by LLMs. However, despite this extensive research, the accessibility of LLM-generated code, which is vital for ensuring equal access to software, remains largely underexplored.

This is particularly crucial in web development, given the growing use of the internet across diverse devices~\cite{henry2014role}, with over four billion people worldwide relying on it~\cite{ourworldindataInternet}. Othman et al. ~\cite{othman2023fostering} investigated using ChatGPT as a tool to enhance the accessibility of existing webpages, while López-Gil et al.\cite{lopez2024turning} explored LLMs' ability to evaluate web accessibility success criteria. Aljedaani et al.~\cite{10.1145/3677846.3677854} examined ChatGPT’s ability to generate accessible code and found that it often produces code with accessibility issues. However, their study did not compare ChatGPT’s performance against human-written code in terms of accessibility, nor did they propose methods for improving the accessibility of LLM-generated code. This leaves a gap in understanding how LLM-generated code compares to human-authored code and what strategies could enhance its accessibility.

In this paper, we aim to fill this gap by comparing the accessibility of LLM-generated code with that of human-authored code for the web and answer the following research questions, among others:

\textbf{RQ1: Do LLMs generate more accessible code than humans?}

We first assess the current state of accessibility violations in code generated by LLMs. To do this, we select ten real-world web projects and regenerate key files affecting the User Interface (UI) of each project using two LLMs: GPT-4o~\cite{gpt4o}, and Qwen2.5-Coder-32B-Instruct-AWQ~\cite{hui2024qwen25codertechnicalreport}. We then analyze the accessibility of these regenerated websites using IBM Equal Access Accessibility Checker~\cite{githubGitHubIBMaequalaccess} and QualWeb Web Accessibility Evaluator~\cite{qualweb}. Finally, we compare accessibility violations in the original implementations of websites with those in their recreated versions using LLM-generated code,

\textbf{RQ2: Do advanced prompting techniques help LLMs generate more accessible code?}

Next, to investigate the LLMs' capability of generating accessible code, we implement three prompting techniques: Zero-Shot, Few-Shot, Self-Criticism~\cite{tan-etal-2023-self}. These techniques are employed in the code generation phase to improve the accessibility of LLM-generated code. We evaluate the accessibility of the websites regenerated using these prompting techniques and compare the accessibility violations.

\textbf{RQ3: Can LLMs generate more accessible code by incorporating accessibility evaluation results into the code generation process?}

To overcome the problem of existing prompting techniques in producing accessible code, where blindly applying accessibility properties or fixes to the code can cause conflicts with the existing code, we introduce \textit{FeedA11y}, a ReAct (Reasoning + Acting) based approach that integrates accessibility evaluation results as feedback for code generation. Our method begins by generating code using an LLM and applying an accessibility evaluation tool to detect violations. The identified issues are used as targeted feedback, enabling the model to refine the code and improve its accessibility iteratively.

In summary, our contributions in this paper are as follows:
\begin{itemize}
    \item Exploring the accessibility status of LLM-generated code in the web domain by using two different LLMs to recreate ten real-world web projects and evaluating the accessibility of the regenerated websites.
    \item Investigating how prompting techniques affect the accessibility of LLM-generated code. 
    \item Proposing \textit{FeedA11y}, a ReAct-based technique to improve the accessibility of LLM-generated code by integrating accessibility evaluation results in the code generation process.
\end{itemize}

\section{Related Works}
\label{sec:rw}

\subsection{Studies on Web Accessibility}
Studies on web accessibility aim to provide access to all users, regardless of disabilities~\cite{campoverde2020empirical, henry2014role}. This includes research on frameworks~\cite{miranda2021web, rodriguez2017framework, angkananon2015technology, heron2013access} designed to support accessible web development and adaptive user interfaces~\cite{raufi2015methods, engel2019svgplott} that adjust to users' varying needs. Additionally, research has focused on design aspects~\cite{ferati2016web, johari2013web, rothberg2019designing, idrobo2017accessibility, friedman2007web} to create inclusive and user-friendly web experiences for individuals with disabilities.

Evaluation techniques and tools for web accessibility have been actively investigated. The most widely recognized guideline is WCAG~\cite{wcag}, an international standard for making the web accessible to people with disabilities. WCAG is structured into principles, guidelines, success criteria, and techniques to address various disabilities. Tools like IBM Equal Access Accessibility Checker~\cite{githubGitHubIBMaequalaccess}, QualWeb~\cite{qualweb}, WAVE~\cite{webaimWAVEAccessibility}, and AXE Accessibility Checker~\cite{dequeAxeAccessibility} automatically evaluate website compliance, identifying issues such as low contrast, missing alt text, and improper heading structures. Researchers have assessed the accessibility of websites, including those in education~\cite{campoverde2020empirical, almeraj2021evaluating, manez2021web} and government~\cite{paul2020accessibility, al2021usability, yi2020web}, highlighting widespread non-compliance and areas for improvement.

\subsection{Large Language Models for Code Generation}
The notable performance of LLMs in code-related tasks has impacted software development process. General purpose LLMs such as GPT-4o~\cite{gpt4o}, DeepSeek-R1~\cite{deepseekai2025deepseekr1incentivizingreasoningcapability}, Llama3~\cite{llama3} have been trained on large corpora, with the purpose of using them for various natural language tasks. These models are not only adept at processing natural language but also demonstrate strong capabilities in code-related tasks. For tasks specifically involving code, specialized LLMs like Qwen2.5-Coder~\cite{hui2024qwen25codertechnicalreport} and Copilot~\cite{copilot} have been trained on programming languages, making them skillful in code-related tasks including code generation~\cite{jiang2024surveylargelanguagemodels, 10298349, ren2023misuse, liu2023codegen4libs, li2022competition, luo2023wizardcoder, yu2024codereval}.

LLMs demonstrate even more significant performance when combined with advanced prompting techniques like Few-Shot~\cite{bareiß2022codegenerationtoolsalmost} and ReAct~\cite{yao2022react}, or when fine-tuned for specific coding tasks~\cite{radford2018improving}. These approaches enhance their ability to tackle a wide range of code-related challenges. Similar to other software engineering tasks, code generation performance has seen significant improvements through the effective use of prompting techniques or fine-tuning methods~\cite{radford2018improving, bareiß2022codegenerationtoolsalmost}.

Given the impressive performance of LLMs in code generation, evaluating the quality of their output is crucial to ensure reliability in software development. While LLMs can produce functionally correct code, studies have shown they fall short of expert-level efficiency~\cite{qiu2024efficient, du2024mercury}. Research on LLM-generated code security has uncovered issues, with 29.6\% of code snippets from GitHub Copilot containing vulnerabilities~\cite{fu2023security}. Additionally, LLMs risk data leakage by unintentionally generating sensitive or proprietary code from their training data~\cite{niu2023codexleaks, yang2023gotcha}, and may produce code violating intellectual property rights~\cite{yu2023codeipprompt}. Further studies have highlighted social and multilingual biases in LLM-generated code~\cite{ling2024evaluating, wang2024exploring}.

In terms of accessibility of LLM-generated code, Aljedaani et al.~\cite{10.1145/3677846.3677854} conducted a study examining ChatGPT's ability to generate accessible web code. Their work involved 88 web developers who prompted ChatGPT to create websites with various characteristics, finding that 84\% of the generated websites contained accessibility issues. While they demonstrated that ChatGPT could fix approximately 70\% of accessibility violations in both its own generated code and third-party open-source projects, their approach was manual and focused solely on ChatGPT. Their study neither compared LLM-generated code against human-written code in terms of accessibility nor proposed systematic methods for improving accessibility during the code generation process. Furthermore, to the best of our knowledge, no prior work has utilized the ReAct approach or incorporated external accessibility evaluation feedback for improving code accessibility. Despite extensive research assessing the quality of LLM-generated code from various perspectives, the accessibility of such code remains largely unexplored, revealing a gap in accessibility research. As interest in accessibility grows and LLM usage in software development increases, examining the accessibility of LLM-generated code becomes essential. To address this gap, we conduct the first empirical study comparing the accessibility of LLM-generated code with human-authored code and propose methods to improve its accessibility through our \textit{FeedA11y} framework.
\section{Methodology}
\label{sec:methodology}

\begin{figure}[htp]
    \centering
    \includegraphics[width=0.7\columnwidth]{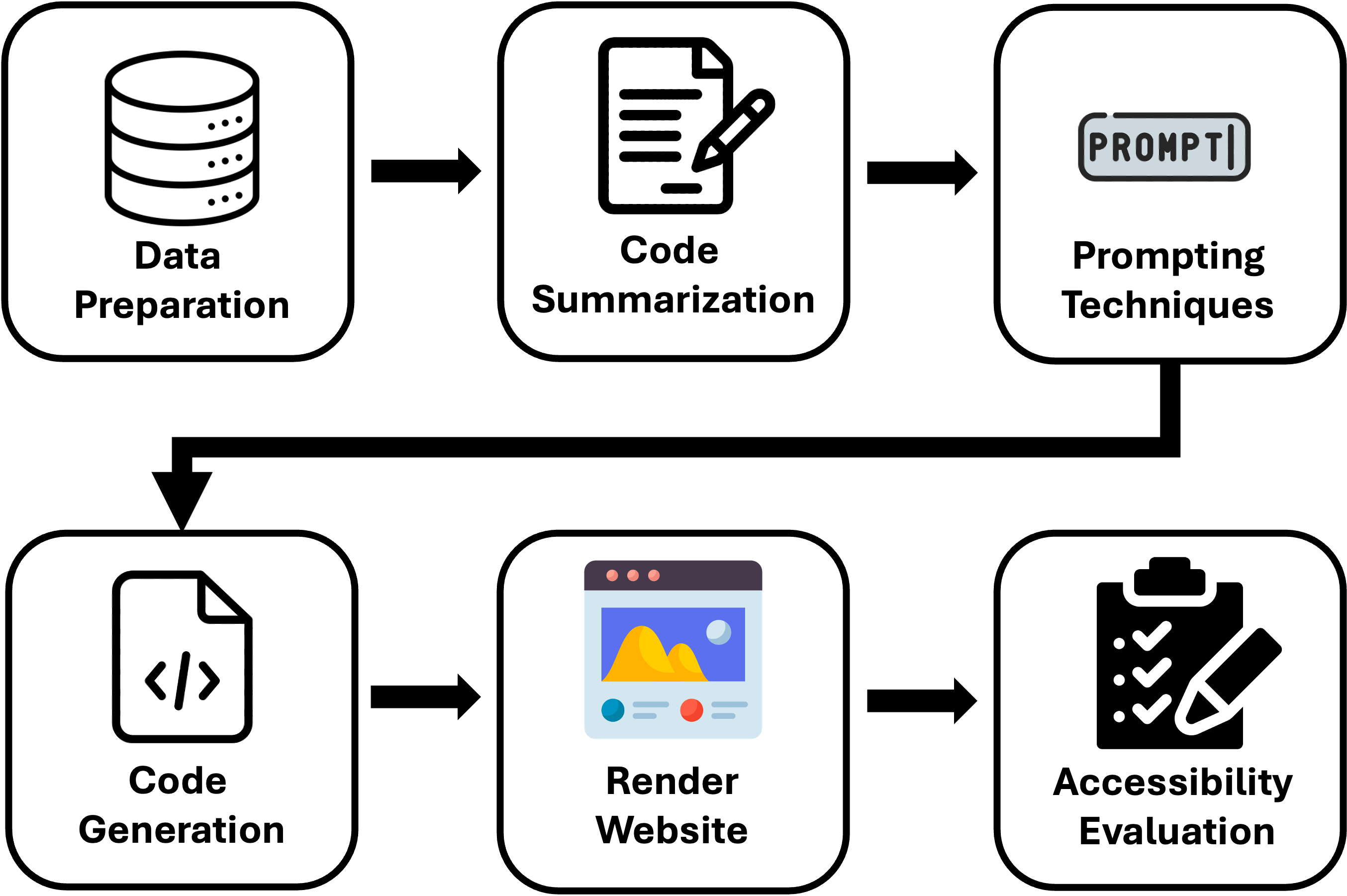}
    \caption{Methodology Overview}
    \label{fig:methodology_overview}
\end{figure}

In this paper, we first investigate the accessibility of LLM-generated web code and then apply advanced prompting techniques to improve LLMs' ability to produce accessible code. Finally, we propose \textit{FeedA11y}, a novel technique that mitigates accessibility violations during code generation by leveraging accessibility evaluation results. This section details our methodology, summarized in Figure~\ref{fig:methodology_overview}.

\subsection{Data Preparation}
\label{subsec:data_preparation}

\begin{table*}
\centering
\caption{Selected Web Projects and File Statistics}
\label{tab:projects_merged}
\footnotesize
\begin{tabular}{l|r|r|r|r|r|r} 
\hline
\multirow{2}{*}{\textbf{Project}} & \multicolumn{3}{c|}{\textbf{GitHub Statistics}} & \multicolumn{3}{c}{\textbf{File Statistics}} \\
\cline{2-7}
 & \textbf{Commits} & \textbf{Stars} & \textbf{Contributors} & \textbf{UI Related Files} & \textbf{Final Size} & \textbf{Sampled Size} \\ 
\hline
\textbf{vuejs/docs}                    & 2,375  & 3.0k  & 619  & 86  & 82  & 39 \\ 
\textbf{django/djangoproject.com}      & 2,885  & 1.9k  & 171  & 87  & 54  & 44 \\ 
\textbf{twbs/bootstrap}                & 22,936 & 172k  & 1,392 & 330 & 217 & 58 \\ 
\textbf{nodejs/nodejs.org}             & 6,106  & 6.4k  & 838  & 103 & 45  & 43 \\ 
\textbf{expressjs/expressjs.com}       & 2,667  & 5.3k  & 540  & 481 & 441 & 60 \\ 
\textbf{flutter/website}               & 7,385  & 2.9k  & 1,093 & 378 & 352 & 58 \\ 
\textbf{postgres/pgweb}                & 2,356  & 77    & 28   & 71  & 67  & 36 \\ 
\textbf{foundation/foundation-sites}   & 17,309 & 29.7k & 1,039 & 100 & 48  & 43 \\ 
\textbf{dart-lang/site-www}            & 4,158  & 979   & 391  & 387 & 370 & 59 \\ 
\textbf{facebook/react-native-website} & 3,691  & 2.0k  & 1,304 & 59  & 40  & 33 \\
\hline
\textbf{Total}                & 71,868 & 224.2k & 7,415 & 2,082 & 1,716 & 473 \\
\hline
\end{tabular}
\end{table*}

Bi et al.~\cite{9609147} found that approximately 70\% of selected GitHub projects reported accessibility issues, with the majority related specifically to UI design elements such as color usage and layout navigation, particularly in the application software domain (72.7\%). Motivated by these findings, we focus our accessible code generation efforts explicitly on UI components. We selected ten active, real-world web projects with interactive UIs, as shown in Table~\ref{tab:projects_merged}. We followed a similar criteria from the prior work~\cite{kalliamvakou2014promises} to avoid toy or non-representative examples: 

\begin{itemize}
    \item Provides an actual website with UI.
    \item Provides a GitHub repository for the website.
    \item The GitHub repository has more than 50 stars.
    \item The GitHub repository has more than 20 contributors.
    \item The latest commit occurred no later than January 2025.
\end{itemize}

To investigate the accessibility of LLM-generated code, we focused on files responsible for rendering UI elements on webpages. We identified UI-related files using two criteria: (1) common UI file extensions (e.g., `.html', `.js', and `.css'), and (2) the presence of HTML tags or UI-related code snippets (e.g., `document.getElementById', `innerHTML', `background:', `font:'). A complete list is available in our replication package~\cite{replicationpackage}. Table~\ref{tab:projects_merged} summarizes the collected files, including the total number of UI-related files, those successfully used for rendering (``Final size''), and the analysis sample (``Sampled size'') selected with 90\% confidence and 10\% margin of error.

\subsection{Code Generation}
With the list of UI-related files obtained in Section~\ref{subsec:data_preparation}, we needed to generate code that could be compared with human-written code for accessibility evaluation. This presented a methodological challenge: we needed to provide specifications to the LLMs, but formal specifications for these files were unavailable. To address this, we employed a two-step approach: first generating code summaries to serve as proxy specifications, then using these summaries to regenerate the code with different LLM instances.

This process aligns with Wei et al.~\cite{wei2019code}, who explored the duality between Code Summarization (CS) and Code Generation (CG), demonstrating that the output of CS can serve as the input for CG, and vice versa. By using separate instances of the same LLM for summarization and generation, we ensured that the LLM responsible for code generation did not have the original code within its context window. This prevented it from simply memorizing and reproducing the source code, forcing it to rely solely on the summarized information for regeneration. Below, we detail each step of this pipeline.

When implementing our code generation pipeline, we needed to determine the optimal granularity for code generation. To regenerate code files, we first attempted a file-level approach, where we generated a summary of the entire file and used it as input to regenerate the code. However, this method resulted in a significant number of files being filtered out during rendering due to functional errors and dependency issues, leading to incorrect or incomplete outputs. To improve code generation accuracy and mitigate rendering issues with reconstructed code, we adopted a block-level approach. 

In our block-level approach, we first parsed the files into blocks based on their structure. We ensured that each code block contained meaningful content by leveraging structural elements. Specifically, for HTML files, we parsed based on structural tags as described in MDN Web Docs~\cite{mozillaStructuringDocuments}. Each block encompassed the content within a high-level structural tag, such as \texttt{<section>}, \texttt{<header>}, \texttt{<nav>}, \texttt{<main>}, or a top-level \texttt{<div>}, ensuring that the segmentation retained meaningful sections of the document. For JavaScript files, we identified functions and classes to create distinct code blocks, and for CSS files, we parsed each declaration block individually. This granular approach allowed us to maintain structural integrity while enabling more accurate regeneration of code components.

To compare whole-file generation and block-level generation objectively, we conducted an empirical study on three web projects (\textit{expressjs/expressjs.com, flutter/website, vuejs/docs}). We evaluated the approaches by analyzing the proportion of UI-related files that successfully passed the rendering phase. On average, block-level generation achieved a pass rate of 93.38\% compared to only 14.98\% for whole-file generation, demonstrating its superiority.

We then generated summaries for each code block by inputting the code content into two LLMs, GPT-4o~\cite{gpt4o}, Qwen2.5-Coder-32B-Instruct-AWQ~\cite{hui2024qwen25codertechnicalreport}, chosen for their notable performance in code-related tasks~\cite{jiang2024surveylargelanguagemodels, liu2024your}. For simplicity, we refer to Qwen2.5-Coder-32B-Instruct-AWQ as Qwen2.5-Coder. For generating summaries, we employed prompt from previous code summarization work~\cite{dhulshette2025hierarchicalrepositorylevelcodesummarization}, instructing LLMs to include function name, inputs, outputs, purpose, workflow, overview, and other factors crucial in understanding the code. To simulate a real-world scenario where developers typically use the default temperature setting of an LLM unless adjustments are necessary, we used the default temperature of 1 for all the LLMs we used in our experiments. To verify the quality and accuracy of the generated summaries, two authors independently conducted a manual inspection to assess whether the summaries comprehensively captured the key characteristics of the code. This process involved two steps. First, both authors independently reviewed the code to understand its functionality. Next, they examined the summary to determine if it accurately reflected the code. A small number of summaries were adjusted if they lacked sufficient information to understand the code. Any discrepancies were then discussed, and through a process of negotiated agreement~\cite{forman2007qualitative}, they reached 100\% consensus.

\begin{table*}
\centering
\caption{Accessibility Guidelines Identified in Analysis Results (AChecker)}
\label{tab:achecker_wcag_rules}
\footnotesize
\resizebox{\textwidth}{!}{%
\begin{tabular}{l|l|l} 
\hline
\multicolumn{1}{c|}{\textbf{Rule ID}}       & \multicolumn{1}{c|}{\textbf{WCAG Technique ID}} & \multicolumn{1}{c}{\textbf{Description}}                                                                \\ 
\hline
\textbf{text\_contrast\_sufficient}         & G18, G145                                       & The contrast ratio of text with its background must meet WCAG AA requirements.                          \\ 
\hline
\textbf{svg\_graphics\_labelled}            & -                                               & A non-decorative SVG element must have an accessible name.                                              \\ 
\hline
\textbf{aria\_hidden\_nontabbable}          & -                                               & A hidden element should not contain any tabbable elements.                                              \\ 
\hline
\textbf{img\_alt\_valid}                    & H37, G94, F38                                   & Images must have accessible names unless they are decorative or redundant.                              \\ 
\hline
\textbf{img\_alt\_redundant}                & H2                                              & The text alternative for an image within a link should not repeat the link text or adjacent link text.  \\ 
\hline
\textbf{input\_label\_exists}               & H44                                             & Each form control must have an associated label.                                                        \\ 
\hline
\textbf{label\_ref\_valid}                  & H44                                             & The 'for' attribute for a label must reference a non-empty, unique 'id' attribute of an input element.  \\ 
\hline
\textbf{a\_text\_purpose}                   & H30                                             & Hyperlinks must have an accessible name for their purpose.                                              \\ 
\hline
\textbf{aria\_child\_tabbable}              & -                                               & UI components must have at least one tabbable descendant for keyboard access.                           \\ 
\hline
\textbf{aria\_complementary\_labelled}      & -                                               & Each element with the "complementary" role must have a label that describes its purpose.                \\ 
\hline
\textbf{aria\_navigation\_label\_unique}    & ARIA6, ARIA13                                   & Each element with the "navigation" role must have a unique label that describes its purpose.            \\ 
\hline
\textbf{aria\_id\_unique}                   & -                                               & The ARIA property must reference a non-empty unique id of an existing element that is visible.          \\ 
\hline
\textbf{aria\_complementary\_label\_unique} & ARIA6, ARIA13                                   & Each element with the "complementary" role must have a unique label that describes its purpose.         \\ 
\hline
\textbf{frame\_title\_exists}               & H64                                             & Inline frames must have a unique, non-empty 'title' attribute.                                          \\ 
\hline
\textbf{table\_headers\_exists}             & H43, H63                                        & Data tables must identify headers.                                                                      \\ 
\hline
\textbf{aria\_banner\_label\_unique}        & ARIA6, ARIA13                                   & Each element with the "banner" role must have a unique label that describes its purpose.                \\ 
\hline
\textbf{aria\_banner\_single}               & -                                               & A page, document, or application should only have one element with the "banner" role.                   \\ 
\hline
\textbf{label\_name\_visible}               & -                                               & Accessible names must match or contain the visible label text.                                          \\ 
\hline
\textbf{aria\_widget\_labelled}             & ARIA4                                           & Interactive components must have a programmatically associated name.                                    \\ 
\hline
\textbf{element\_scrollable\_tabbable}      & G202                                            & Scrollable elements should be tabbable or contain tabbable content.                                     \\ 
\hline
\textbf{html\_lang\_exists}                 & H57                                             & The page must identify the default language of the document with a 'lang' attribute.                    \\ 
\hline
\textbf{input\_label\_after}                & -                                               & An input element must be labeled.                                                                       \\ 
\hline
\textbf{label\_content\_exists}             & -                                               & A label element must have non-empty text or an element with an accessible name.                         \\ 
\hline
\textbf{table\_scope\_valid}                & H63                                             & The scope attribute must be used correctly to associate table headers and data cells.                   \\ 
\hline
\textbf{aria\_contentinfo\_label\_unique}   & ARIA6, ARIA13                                   & Each element with the "contentinfo" role must have a unique label that describes its purpose.           \\ 
\hline
\textbf{aria\_contentinfo\_single}          & -                                               & A page, document, or application should only have one element with the "contentinfo" role.              \\ 
\hline
\textbf{aria\_main\_label\_unique}          & ARIA6, ARIA13                                   & Each element with the "main" role must have a unique label that describes its purpose.                  \\ 
\hline
\textbf{aria\_region\_label\_unique}        & ARIA6, ARIA13                                   & Each element with the "region" role must have a unique label that describes its purpose.                \\ 
\hline
\textbf{aria\_role\_valid}                  & ARIA4                                           & Elements must have valid roles per the ARIA specification.                                              \\ 
\hline
\textbf{combobox\_popup\_reference}         & -                                               & A combobox must reference a valid popup element.                                                        \\ 
\hline
\textbf{element\_orientation\_unlocked}     & -                                               & Content must not restrict its orientation to a single display orientation.                              \\ 
\hline
\textbf{page\_title\_exists}                & G88                                             & The page should have a title that correctly identifies the subject of the page.                         \\ 
\hline
\textbf{skip\_main\_exists}                 & G1                                              & Pages must provide a way to skip directly to the main content.                                          \\ 
\hline
\textbf{table\_headers\_related}            & H43, H63                                        & Table headers must be related to their corresponding data cells.                                        \\
\hline
\end{tabular}%
}
\end{table*}

\begin{table*}[ht]
\centering
\caption{Accessibility Guidelines Identified in Analysis Results (QualWeb)}
\label{tab:qualweb_wcag_rules}
\footnotesize
\resizebox{\textwidth}{!}{%
\begin{tabular}{l|l|l} 
\hline
\multicolumn{1}{c|}{\textbf{Rule ID}} & \multicolumn{1}{c|}{\textbf{WCAG Technique ID}} & \multicolumn{1}{c}{\textbf{Description}} \\ 
\hline
\textbf{AltFailure} & F30 & Failure of Success Criterion 1.1.1 and 1.2.1 due to using text alternatives that are not alternatives \\ 
\hline
\textbf{CaptionDataTbl} & H39 & Using caption elements to associate data table captions with data tables \\ 
\hline
\textbf{ColorContrastFail} & F24 & Failure of Success Criterion 1.4.3, 1.4.6 and 1.4.8 due to specifying foreground colors without specifying background colors or vice versa \\ 
\hline
\textbf{CombineAdj} & H2 & Combining adjacent image and text links for the same resource \\
\hline
\textbf{FocusRemoveFail} & F55 & Failure of Success Criteria 2.1.1, 2.4.7, and 3.2.1 due to using script to remove focus when focus is received \\ 
\hline
\textbf{FontSizeCSS} & C12, C13, C14 & Using percent, em, names for font sizes \\ 
\hline
\textbf{HeadingsOrg} & G141 & Organizing a page using headings \\ 
\hline
\textbf{IdHeadersDataTbl} & H43 & Using id and headers attributes to associate data cells with header cells in data tables \\ 
\hline
\textbf{ImgLinkFail} & F89 & Failure of Success Criteria 2.4.4, 2.4.9, and 4.1.2 due to not providing an accessible name for an image which is the only content in a link \\ 
\hline
\textbf{LabelPos} & G162 & Positioning labels to maximize predictability of relationships \\ 
\hline
\textbf{LayoutTblFail} & F46 & Failure of Success Criterion 1.3.1 due to using th elements, caption elements, or non-empty summary attributes in layout tables \\ 
\hline
\textbf{LinkTitleAttr} & H33 & Supplementing link text with the title attribute \\ 
\hline
\textbf{ListLinkGroups} & H48 & Using ol, ul, and dl for lists or groups of links \\ 
\hline
\textbf{ScopeDataTbl} & H63 & Using the scope attribute to associate header cells and data cells in data tables \\ 
\hline
\textbf{SkipToMain} & G1 & Adding a link at the top of each page that goes directly to the main content area \\ 
\hline
\textbf{SubmitBtn} & H32 & Providing submit buttons \\ 
\hline
\textbf{TblMarkup} & H51 & Using table markup to present tabular information \\
\hline

\end{tabular}%
}
\end{table*}

After generating summaries with LLMs, we used these summaries as inputs to regenerate code, incorporating them into prompts fed to separate LLM instances. To prevent leakage, we ensured that the LLM responsible for generating code had no access to the original code used for summarization. Prompts instructed the LLMs to generate code solely based on the summaries, with examples provided in Section~\ref{subsec:prompting_techniques}. Notably, for RQ1, we did not mention accessibility in any prompts, as our goal was to assess whether LLMs can inherently produce accessible code without explicit guidance—a process we refer to as \textit{Naive Code Generation}.

Using the regenerated code, we rendered webpages, specifically focusing on the ten selected URLs for each web project mentioned in Table \ref{tab:projects_merged}. To ensure that the rendered websites using regenerated code maintain the same structure as the original versions, two authors manually inspected the UI differences and confirmed that the structure remained consistent across all URLs. A webpage was considered consistent if its main UI components remained in place, no content was altered, and the visual layout did not exhibit significant deviations. Minor variations, such as slight differences in padding, font weight, or margin spacing, were deemed acceptable as long as they did not impact readability or usability. 

We conducted this verification process manually, as no existing automated tools met our specific requirements for comparing visual and structural consistency between original and regenerated webpages. So we adopted a qualitative approach instead. Two authors collaboratively analyzed the results, discussing any discrepancies that arose. Through a process of negotiated agreement~\cite{forman2007qualitative}, they resolved differences and achieved 100\% consensus.

\subsection{Accessibility Evaluation}
\label{subsec:accessibility_evaluation}


For the accessibility evaluation, we employed two tools: the IBM Equal Access Accessibility Checker (AChecker)~\cite{githubGitHubIBMaequalaccess} and the QualWeb Accessibility Tool~\cite{qualweb}. We selected AChecker and QualWeb because they have been shown to detect the highest number of accessibility issues compared to other tools~\cite{tafreshipour2024ma11y}. We evaluated the code against the WCAG 2.1 guidelines~\cite{wcag}, as this was the most recent version supported by both AChecker and QualWeb. The accessibility guidelines identified in our evaluation results are organized in Table~\ref{tab:achecker_wcag_rules} for AChecker and Table~\ref{tab:qualweb_wcag_rules} for QualWeb. For clarity, we generated a rule ID for each QualWeb rule to represent the content of the accessibility guide description, as the original rule names were too lengthy to display in the paper. For AChecker, we used the original rule IDs. In Table~\ref{tab:achecker_wcag_rules}, some cells in the WCAG Technique ID column contain dashes (-) because certain accessibility rules implemented by AChecker are not directly mapped to specific WCAG techniques. These rules are based on general accessibility best practices or are implementation-specific checks that AChecker performs without referencing a particular WCAG technique.

We exhaustively evaluated all extracted URLs for each web project. Understanding the relationship between URLs and files is crucial: URLs correspond to the actual rendered web pages, while multiple files could be utilized to generate a single web page. For each of these URLs, we ran AChecker and QualWeb to evaluate the accessibility of the rendered web page.

The evaluation results from AChecker classify accessibility issues into several levels: \textit{violation}, \textit{potential violation}, \textit{recommendation}, \textit{potential recommendation}, and \textit{manual}. A \textit{violation} denotes an accessibility issue that fails to meet the WCAG standards and must be addressed immediately to ensure compliance. A \textit{potential violation} refers to an issue that may be non-compliant but cannot be confirmed automatically by the checker. A \textit{recommendation} is a suggested improvement for accessibility, rather than a requirement. A \textit{potential recommendation} identifies areas that might benefit from enhancement and require further review. The manual category indicates issues that require human judgment for proper evaluation, as they cannot be fully detected automatically. For our experiment, we focused on \textit{violations} in AChecker's results, as they represent the most critical issues requiring immediate attention.

QualWeb categorizes the evaluation results using different levels: \textit{Passed}, \textit{Warnings}, \textit{Failed}, \textit{Inapplicable}. A \textit{Passed} result indicates that the inspected element does not contain any accessibility issues. \textit{Warnings} signify that the element has potential issues that could affect accessibility but are not immediately critical. \textit{Failed} denotes that the element causes accessibility failures and must be addressed promptly. \textit{Inapplicable} means that the element was not relevant for evaluation. Similar to AChecker, our focus is on the \textit{Failed} category in QualWeb's results. For consistency throughout the paper, we will refer to the \textit{Failed} category as \textit{violation}.

\begin{figure}[t]
\centering
\begin{tikzpicture}[
  scale=0.9, transform shape, 
  box/.style={
    rectangle,
    draw=black!70,
    fill=gray!10,
    rounded corners=3pt,
    text width=0.9\columnwidth,
    align=justify,
    font=\small,
    line width=0.6pt,
    inner sep=6pt
  },
  highlightbox/.style={
    rectangle,
    fill=yellow!30,
    text width=0.88\columnwidth,
    inner sep=6pt,
    align=justify
  },
  arrow/.style={
    ->,
    >=stealth,
    thick,
    black!70
  }
]
\node[box] (naive) at (0,0) {
    \textbf{Naive Prompt}\\
    Act as a software developer. Write code from a file description covering all necessary aspects.\\
    \textbf{Summary:} [summary]
};
\node[box] (zero) at (0,-1.8) {
    \textbf{Zero-Shot} - Naive +
    \hl{Make code compliant with WCAG accessibility rules. Avoid any violations.}
};
\node[box] (few) at (0,-3.6) {
    \textbf{Few-Shot} - Naive + Zero-Shot +
    
    \hl{Detailed accessibility rules with correct/incorrect examples for each rule.\\
    \textit{Rule structure:}[name, description, correct example, counter example]}
};
\node[box] (self) at (0,-5.8) {
    \textbf{Self-Criticism}\\
    \textit{Generator:} Naive + Zero-Shot\\
    \textit{Reviewer:}
    \hl{Review code for WCAG compliance. Return unchanged if compliant, fix issues if not.}
};
\draw[arrow] (naive) -- (zero);
\draw[arrow] (zero) -- (few);
\draw[arrow] (zero.west) -- ++(-0.5,0) |- (self.west);
\end{tikzpicture}
\caption{Progressive Enhancement of Prompts with Accessibility Instructions}
\label{fig:prompts_condensed}
\end{figure}
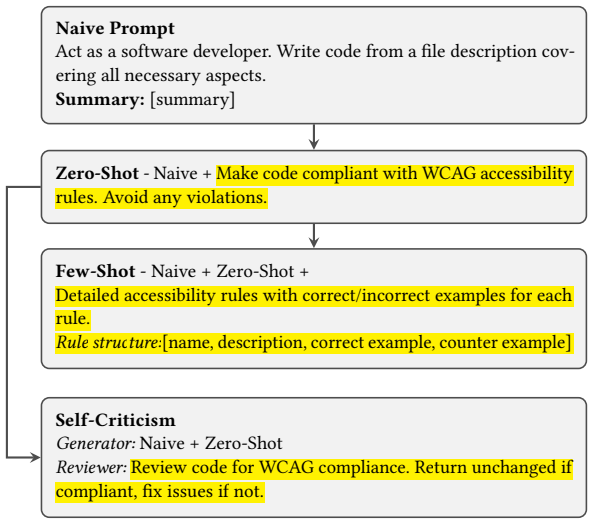

As used in previous work~\cite{10.1145/3377811.3380392}, we used inaccessibility rate as our evaluation metric, which is calculated by dividing the number of elements with accessibility violations over the total number of elements that are prone to accessibility violations. We report the inaccessibility rate for each project, as well as the average inaccessibility rate across all projects for both the LLMs and human evaluations in Section~\ref{subsec:rq1}.

We also examined the types of accessibility violations identified in the evaluation results. Both AChecker and QualWeb categorize violations based on WCAG 2.1 guidelines. To understand how different prompting techniques impact specific types of accessibility violations, we analyzed the inaccessibility rate of elements with each violation type across configurations. To clearly illustrate the differences in the occurrence of each violation, we present the percentage change in the inaccessibility rate for each violation compared to human-written code. Additionally, we classified violations by severity and aligned them with the four main WCAG principles—Perceivable, Operable, Understandable, and Robust—to gain deeper insights into the nature and impact of these violations. 

\subsection{Prompting Techniques}
\label{subsec:prompting_techniques}

\begin{table}
\centering
\caption{Prompting Techniques Comparison}
\label{tab:pt_comparison}
\footnotesize
\setlength{\tabcolsep}{0.55em}
\begin{tabular*}{\columnwidth}{@{\extracolsep{\fill}}l|>{\centering\arraybackslash}p{0.8cm}|>{\centering\arraybackslash}p{0.8cm}|>{\centering\arraybackslash}p{0.8cm}|>{\centering\arraybackslash}p{0.8cm}|>{\centering\arraybackslash}p{0.8cm}@{}} 
\hline
\textbf{Feature} & \textbf{N} & \textbf{Z} & \textbf{F} & \textbf{S} & \textbf{FA} \\
\hline
Code Summary & \checkmark & \checkmark & \checkmark & \checkmark & \checkmark \\
A11y Instructions & \texttimes & \checkmark & \checkmark & \checkmark & \checkmark \\
Guidelines & \texttimes & \texttimes & \checkmark & \texttimes & \checkmark \\
Code Examples & \texttimes & \texttimes & \checkmark & \texttimes & \checkmark \\
Test Rules & \texttimes & \texttimes & \texttimes & \texttimes & \checkmark \\
Style Properties & \texttimes & \texttimes & \texttimes & \texttimes & \checkmark \\
Multiple Rounds & \texttimes & \texttimes & \texttimes & \texttimes & \checkmark \\
Multiple Agents & \texttimes & \texttimes & \texttimes & \checkmark & \checkmark \\
\hline
\end{tabular*}
\footnotesize \textbf{N}: Naive Generation, \textbf{Z}: Zero-Shot, \textbf{F}: Few-Shot, \textbf{S}: Self-Criticism, \textbf{FA}: FeedA11y
\end{table}

Prior studies have demonstrated that various prompting techniques, such as Zero-Shot~\cite{kojima2022large}, Few-Shot~\cite{lin2021few}, Self-Criticism~\cite{tan-etal-2023-self} can enhance the ability of LLMs to generate more accurate and reliable outputs. Accordingly, we applied these prompting techniques to reduce the number of accessibility violations in the generated code. Specifically, we used Zero-Shot, Few-Shot, Self-Criticism. We followed the established best practices~\cite{awesomeprompt,ibm} to design our prompt.

We present the prompts used in our experiment in figure~\ref{fig:prompts_condensed}. The prompts highlighted in color represent the instructions relevant to accessibility. In each prompting technique we employed, we instructed the LLMs to act as if they were a software developer. For Zero-Shot prompting, we explicitly instructed the model to consider accessibility guidelines when generating code. The key distinction between Zero-Shot and \textit{Naive Code Generation} is that Zero-Shot specifically emphasizes accessibility, whereas \textit{Naive Code Generation} does not address accessibility in its prompts.

In Few-Shot prompting, we enhanced the prompt with additional examples to help the LLM generate more accessible code. We listed the types of accessibility violations that occurred in RQ1 for AChecker and QualWeb. For each of these violation types, the examples were sourced from the rule sets provided by accessibility evaluation tools~\cite{ibmAccessibilityRequirements, qualweb}. The examples included correct code that adheres to accessibility requirements, as well as the mutation of the correct code with accessibility issues~\cite{tafreshipour2024ma11y}. We manually verified that these modifications correctly triggered specific violations. Both the correct examples and counterexamples were included in the prompt to guide the LLM in generating compliant code. We ultimately provided correct examples and counterexamples for each of the 34 violation types identified by AChecker, and 16 violation types identified by QualWeb.

For Self-Criticism, we first instructed the LLMs to generate code with adherence to accessibility guidelines in mind. After generating the code, we prompted the LLMs again with their own output, asking them to review and assess whether it met the accessibility guidelines. If the review identified any accessibility issues, the LLMs were asked to return revised code addressing those issues. If no issues were found, the LLMs were instructed to return the code as originally generated. We went through a single round of code generation and review for each code.

To provide a clear understanding of the features associated with each prompting technique, we present Table~\ref{tab:pt_comparison}. Code Summary refers to an initial summary of the code provided in the prompt, which is applied across all prompting techniques. Instructions on Accessibility indicate explicit instructions in the prompt that direct the model to generate accessible code as highlighted in Figure~\ref{fig:prompts_condensed}. Guideline Descriptions consist of accessibility guidelines sourced from accessibility evaluators. Code Examples include actual code snippets for each violation, featuring both a correct implementation and a mutated version containing the violation. Test Rules are WCAG-based testing rules that provide systematic guidance on detecting specific accessibility violations. Lastly, Multiple Rounds and Multiple Agents pertain to the architectural design of the prompting techniques.

After generating code using each prompting technique and conducting an accessibility evaluation, we analyzed the types of violations that occurred to assess the effectiveness of each technique in improving accessibility. 

Despite our expectations that explicit accessibility instructions would improve accessibility, our analysis revealed that the existing prompting techniques had limitations. Specifically, we observed that when accessibility attributes were introduced through these techniques, they sometimes conflicted with the existing codebase structure, leading to new accessibility issues rather than resolving them. These findings motivated us to develop a more sophisticated approach that could integrate accessibility evaluation feedback directly into the generation process, which we introduce in the following section as \textit{FeedA11y.}

\subsection{FeedA11y}
\label{subsec:feeda11y}

To address the limitations of existing prompting techniques for generating accessible code, we propose \textit{FeedA11y}, a ReAct (Reasoning + Acting)~\cite{yao2022react}-based method that integrates accessibility evaluation reports as feedback into the generation process. By leveraging ReAct framework, \textit{FeedA11y} guides the LLM to systematically reason through accessibility violations, apply targeted fixes, and verify improvements before producing the final code.

\begin{figure}[htp]
    \centering
    \includegraphics[width=0.95\columnwidth]{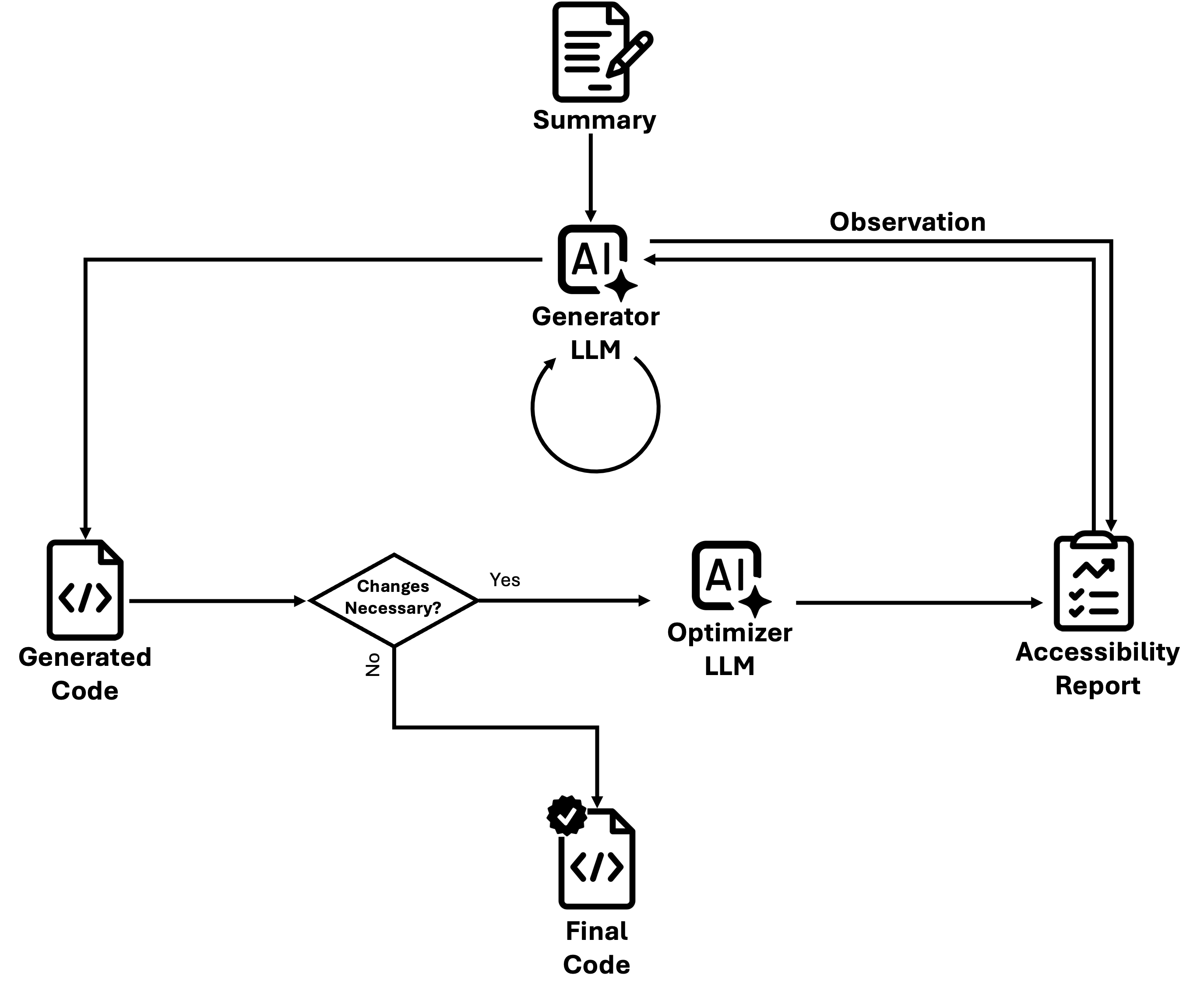}
    \caption{FeedA11y Overview}
    \label{fig:feeda11y}
\end{figure}

The overall structure of \textit{FeedA11y} is represented in Fig~\ref{fig:feeda11y}. The process begins with inputting a summary into the Generator LLM to generate an initial version of the code. Notably, this initial code is generated without any specific instructions for accessibility improvements. This approach differs from including accessibility instructions in the initial prompt. Our analysis of prompting techniques showed that generic accessibility instructions often lead to over-application of accessibility attributes in inappropriate contexts, creating new issues. By starting with a ``clean'' generation and then using precise, targeted feedback from actual accessibility evaluations, FeedA11y can apply specific fixes exactly where needed rather than broadly applying accessibility patterns that might conflict with the existing code structure.

After generating the initial code, we use an Optimizer LLM, a separate model from the Generator LLM, to evaluate the code's accessibility. To ensure an accurate evaluation, we provide the Optimizer LLM with several key pieces of context information: the generated code, accessibility guidelines from the QualWeb \textbf{or} AChecker evaluation tool, style properties of the elements, and test rules sourced from WCAG success criteria. Notably, we use guidelines from different accessibility evaluation tools for both optimization and evaluation, ensuring they are not applied in the same process. For example, QualWeb rules are used for optimization while AChecker rules are used for evaluation, and vice versa. This separation prevents the same tool from being used for both optimization and evaluation, ensuring a more objective assessment. To incorporate style properties, we extracted element styles by parsing the HTML and CSS files using the jsdom and css libraries. This extraction process creates mappings between HTML elements (identified by their id, class, and tag) and their corresponding CSS declarations, allowing us to determine the exact styling applied to each element in the rendered page. Additionally, the test rules sourced from WCAG success criteria define the specific accessibility violations to detect. We include all this contextual information in the prompt to instruct the LLM to generate a comprehensive accessibility report.

Once the accessibility report is generated, we construct a structured prompt following the ReAct methodology. LLMs are instructed to reason through the problem using a "Thought" section, where it analyzes the violation, considers the necessary modifications, and ensures compliance with the report. The LLMs were instructed to focus only on the accessibility violations specified in the report, without applying additional fixes beyond those identified. This approach was adopted after observing that previous prompting techniques, which broadly instructed LLMs to generate accessible code without a target focus, often led to counterproductive results. The "Action" step follows, where it applies appropriate modifications to the affected snippets while preserving the overall structure and functionality of the code. Finally, in the "Observation" phase, it evaluates whether the applied changes effectively resolved the violation. After obtaining the revised code, we update the code by replacing only the affected parts while keeping the rest of the document intact. We repeated this iterative process until either no further modifications were made between consecutive rounds or we reached a maximum of three refinement rounds. On average, 2.4 iterations were needed to reach convergence, with 78\% of cases requiring two or more rounds to resolve all addressable violations.

\section{Results}
\label{sec:results}

We now study the following research questions: 
\begin{enumerate}[label={}]
    \item [RQ1:] Do LLMs generate more accessible code than humans?
    \item [RQ2:] Do advanced prompting techniques help LLMs generate more accessible code?
    \item [RQ3:] Can LLMs generate more accessible code by incorporating accessibility evaluation results into the code generation process?
    \item [RQ4:] To what extent does data leakage impact the effectiveness of \textit{FeedA11y} in generating accessible code?
    \item [RQ5:] What design choices in \textit{FeedA11y} have the most significant impact on the accessibility of generated code?
\end{enumerate}


\begin{table*}
\centering
\caption{Inaccessibility Rate Computed Using AChecker for GPT-4o and Qwen2.5-Coder}
\label{tab:achecker_results_combined}
\footnotesize
\begin{tabular*}{\textwidth}{@{\extracolsep{\fill}}l|c|cc|cc|cc|cc|cc@{}}
\hline
\multirow{2}{*}{\textbf{Project}} & \multirow{2}{*}{\textbf{H}} & \multicolumn{2}{c|}{\textbf{Naive}} & \multicolumn{2}{c|}{\textbf{Zero-Shot}} & \multicolumn{2}{c|}{\textbf{Few-Shot}} & \multicolumn{2}{c|}{\textbf{Self-Criticism}} & \multicolumn{2}{c}{\textbf{FeedA11y}} \\
\cline{3-12}
 & & \textbf{G} & \textbf{Q} & \textbf{G} & \textbf{Q} & \textbf{G} & \textbf{Q} & \textbf{G} & \textbf{Q} & \textbf{G} & \textbf{Q} \\
\hline
\textbf{vuejs/docs} & 0.488 & 0.475 & 0.496 & 0.500 & 0.496 & 0.488 & 0.477 & 0.475 & 0.453 & 0.475 & 0.489 \\
\textbf{django/djangoproject.com} & 0.894 & 0.771 & 0.840 & 0.871 & 0.840 & 0.719 & 0.671 & 0.774 & 0.840 & 0.399 & 0.651 \\
\textbf{twbs/bootstrap} & 0.108 & 0.153 & 0.323 & 0.128 & 0.591 & 0.355 & 0.674 & 0.318 & 0.590 & 0.185 & 0.271 \\
\textbf{nodejs/nodejs.org} & 0.562 & 0.574 & 0.526 & 0.580 & 0.539 & 0.511 & 0.476 & 0.169 & 0.528 & 0.500 & 0.463 \\
\textbf{expressjs/expressjs.com} & 0.368 & 0.293 & 0.219 & 0.293 & 0.223 & 0.315 & 0.434 & 0.302 & 0.286 & 0.205 & 0.221 \\
\textbf{flutter/website} & 0.391 & 0.170 & 0.196 & 0.630 & 0.314 & 0.598 & 0.680 & 0.219 & 0.547 & 0.152 & 0.159 \\
\textbf{postgres/pgweb} & 0.055 & 0.069 & 0.069 & 0.065 & 0.045 & 0.271 & 0.076 & 0.749 & 0.140 & 0.030 & 0.067 \\
\textbf{foundation/foundation-sites} & 0.823 & 0.394 & 0.385 & 0.432 & 0.319 & 0.412 & 0.509 & 0.391 & 0.376 & 0.400 & 0.346 \\
\textbf{dart-lang/site-www} & 0.310 & 0.338 & 0.254 & 0.234 & 0.215 & 0.654 & 0.286 & 0.172 & 0.345 & 0.343 & 0.164 \\
\textbf{facebook/react-native-website} & 0.255 & 0.229 & 0.169 & 0.126 & 0.138 & 0.169 & 0.163 & 0.238 & 0.123 & 0.228 & 0.168 \\
\hline
\textbf{AVG} & 0.425 & 0.347 & 0.348 & 0.386 & 0.372 & 0.449 & 0.444 & 0.381 & 0.423 & \hl{\textbf{0.292}} & \hl{\textbf{0.300}} \\
\hline
\end{tabular*}

\footnotesize \textbf{H}: Human, \textbf{G}: GPT-4o, \textbf{Q}: Qwen2.5-Coder, \textbf{Naive}: Naive Code Generation
\end{table*}
\begin{table*}
\centering
\caption{Inaccessibility Rate Computed Using QualWeb for GPT-4o and Qwen2.5-Coder}
\label{tab:qualweb_results_combined}
\footnotesize
\begin{tabular*}{\textwidth}{@{\extracolsep{\fill}}l|c|cc|cc|cc|cc|cc@{}}
\hline
\multirow{2}{*}{\textbf{Project}} & \multirow{2}{*}{\textbf{H}} & \multicolumn{2}{c|}{\textbf{Naive}} & \multicolumn{2}{c|}{\textbf{Zero-Shot}} & \multicolumn{2}{c|}{\textbf{Few-Shot}} & \multicolumn{2}{c|}{\textbf{Self-Criticism}} & \multicolumn{2}{c}{\textbf{FeedA11y}} \\
\cline{3-12}
 & & \textbf{G} & \textbf{Q} & \textbf{G} & \textbf{Q} & \textbf{G} & \textbf{Q} & \textbf{G} & \textbf{Q} & \textbf{G} & \textbf{Q} \\
\hline
\textbf{vuejs/docs} & 0.166 & 0.112 & 0.139 & 0.152 & 0.163 & 0.174 & 0.156 & 0.251 & 0.141 & 0.163 & 0.163 \\
\textbf{django/djangoproject.com} & 0.021 & 0.028 & 0.031 & 0.024 & 0.040 & 0.055 & 0.044 & 0.034 & 0.054 & 0.025 & 0.042 \\
\textbf{twbs/bootstrap} & 0.061 & 0.070 & 0.062 & 0.070 & 0.069 & 0.082 & 0.077 & 0.082 & 0.069 & 0.082 & 0.065 \\
\textbf{nodejs/nodejs.org} & 0.021 & 0.018 & 0.025 & 0.041 & 0.013 & 0.026 & 0.023 & 0.022 & 0.092 & 0.025 & 0.019 \\
\textbf{expressjs/expressjs.com} & 0.144 & 0.115 & 0.172 & 0.139 & 0.149 & 0.142 & 0.136 & 0.110 & 0.210 & 0.169 & 0.169 \\
\textbf{flutter/website} & 0.179 & 0.028 & 0.027 & 0.039 & 0.032 & 0.028 & 0.036 & 0.028 & 0.029 & 0.031 & 0.031 \\
\textbf{postgres/pgweb} & 0.057 & 0.088 & 0.091 & 0.071 & 0.084 & 0.060 & 0.056 & 0.073 & 0.091 & 0.044 & 0.067 \\
\textbf{foundation/foundation-sites} & 0.266 & 0.261 & 0.235 & 0.263 & 0.254 & 0.245 & 0.263 & 0.249 & 0.237 & 0.250 & 0.192 \\
\textbf{dart-lang/site-www} & 0.210 & 0.236 & 0.227 & 0.218 & 0.256 & 0.263 & 0.209 & 0.209 & 0.205 & 0.223 & 0.223 \\
\textbf{facebook/react-native-website} & 0.121 & 0.097 & 0.121 & 0.098 & 0.153 & 0.102 & 0.108 & 0.067 & 0.087 & 0.104 & 0.101 \\
\hline
\textbf{AVG} & 0.125 & \textbf{0.105} & 0.113 & 0.112 & 0.121 & 0.118 & 0.111 & 0.112 & 0.122 & 0.112 & \hl{\textbf{0.107}} \\
\hline
\end{tabular*}

\footnotesize \textbf{H}: Human, \textbf{G}: GPT-4o, \textbf{Q}: Qwen2.5-Coder, \textbf{Naive}: Naive Code Generation
\end{table*}

\subsection{RQ1: Do LLMs generate more accessible code than humans?}
\label{subsec:rq1}

We first investigated the accessibility status of human-written and LLM-generated code using the two accessibility evaluation tools, AChecker and QualWeb, and compared the inaccessibility rate. We also analyzed the type of violations that occurred by looking into the inaccessibility rate of elements with each violation. We marked the best performing technique in Table~\ref{tab:achecker_results_combined} and Table~\ref{tab:qualweb_results_combined} as bold, and the cases where FeedA11y achieved the best performance highlighted.

\subsubsection{Evaluation with AChecker.}
Table~\ref{tab:achecker_results_combined} shows that \textit{Naive Code Generation} by GPT-4o and Qwen2.5-Coder produces more accessible code than human-written code, and with lower inaccessibility rates (0.347 and 0.348 vs. 0.425). To better understand accessibility issues, we categorized violations by type, finding that \textit{Contrast Issues} (\textit{text\_contrast\_sufficient}) dominate, comprising 80\% of all violations.

Compared to human-written code, GPT-4o significantly reduced \textit{Contrast Issues} by nearly 49\% and \textit{Alternative Text Issues} such as \textit{img\_alt\_valid} and \textit{img\_alt\_redundant} by over 70\%. It also effectively addressed \textit{Form Labeling Issues} such as \textit{input\_label\_after} and \textit{input\_label\_before}, showing strong performance on common accessibility concerns. However, GPT-4o introduced more violations in areas like ARIA-related attributes, including \textit{Graphics Labeling} (\textit{svg\_graphics\_labelled}), \textit{Form Label Position Issues}, \textit{Link Text Clarity Issues} (\textit{a\_text\_purpose}), and \textit{Frame Accessibility Issues} (\textit{frame\_title\_exists}). These findings suggest GPT-4o handles basic accessibility well but struggles with more complex ARIA implementations.

Qwen2.5-Coder showed similar improvements to GPT-4o in reducing \textit{Contrast} and \textit{Alternative Text Issues}. Unlike GPT-4o, it better handled \textit{ARIA ID Uniqueness Issues} such as \textit{aria\_id\_unique} but struggled with \textit{Label Visibility Issues} such as \textit{label\_name\_visible}, which GPT-4o managed well. Both models faced similar challenges with ARIA-related violations, indicating that complex ARIA handling remains difficult for LLMs.

\subsubsection{Evaluation with QualWeb.}

Similarly, in the QualWeb evaluation, \textit{Naive Code Generation} using both LLMs outperformed human-written code. GPT-4o and Qwen2.5-Coder achieved lower inaccessibility rates (0.105 and 0.113, respectively) compared to humans (0.125). When analyzing types of accessibility violations, both LLMs  notably reduced \textit{ColorContrastFail} violations by ~58\% and improved \textit{SkipToMain} issues, with Qwen2.5-Coder performing especially well. However, both struggled with \textit{FontSizeCSS}, \textit{ScopeDataTbl}, and \textit{HeadingsOrg}. Qwen2.5-Coder also had additional issues with \textit{LinkTitleAttr} and \textit{FocusRemoveFail}.

\begin{tcolorbox}[colback=black!5!white,colframe=gray!75!black,fonttitle=\bfseries,title=Finding 1]
\textbf{LLMs excel at addressing basic accessibility requirements but struggle with complex accessibility requirements, particularly ARIA-related attributes, performing worse than human developers.}
\end{tcolorbox}

\subsubsection*{\underline{Discussion and Implications.}}
The superior performance of LLMs, particularly in basic accessibility concerns such as contrast and alternative text issues, suggests a promising opportunity for automation and efficiency gains in web development workflows. These models could be effectively employed to handle routine accessibility corrections, thereby allowing human developers to focus their attention on more complex accessibility considerations.

Nevertheless, the evident difficulties in managing intricate ARIA attributes highlight an essential limitation of current LLMs—the lack of nuanced understanding of complex semantic requirements. This reinforces the importance of maintaining human oversight and suggests potential improvements in LLM training or prompt engineering specifically targeted at complex ARIA semantics. Practitioners and researchers should leverage these insights to prioritize complementary approaches, blending AI-generated drafts with expert validation, ensuring high accessibility standards without distorting the original meaning or structure of the web content.

\subsection{RQ2: Do advanced prompting techniques help LLMs generate more accessible code?}
\label{subsec:rq2}

To mitigate accessibility violations in LLM-generated code, we applied different prompting techniques during the code generation phase. We present the inaccessibility rates for these techniques as evaluated by AChecker  (Table~\ref{tab:achecker_results_combined}) and QualWeb (Table~\ref{tab:qualweb_results_combined}).

\subsubsection{Evaluation with AChecker.}
In the AChecker evaluation, Zero-Shot and Self-Criticism improved accessibility over human-written code (0.425 inaccessibility rate) but underperformed compared to \textit{Naive Code Generation}. For GPT-4o, Zero-Shot achieved 0.386, and Self-Criticism 0.381, both better than humans but worse than \textit{Naive Code Generation} (0.347). Few-Shot performed the worst, with a 0.449 inaccessibility rate, exceeding all other methods.

Qwen2.5-Coder followed a similar trend. Zero-Shot (0.372) and Self-Criticism (0.423) slightly outperformed human-written code but remained less effective than \textit{Naive Code Generation} (0.348), reinforcing the pattern seen with GPT-4o.

When comparing the three prompting techniques used with GPT-4o, Few-Shot exhibited the highest inaccessibility rate (0.449), followed by Zero-Shot (0.386) and Self-Criticism (0.381). For Qwen2.5-Coder, Few-Shot again had the highest inaccessibility rate (0.444), followed by Self-Criticism (0.423) and Zero-Shot (0.372).

\subsubsection{Evaluation with QualWeb.}

When evaluated using QualWeb, all three prompting techniques—Zero-Shot, Few-Shot, and Self-Criticism—resulted in code that was more accessible than human-written code, achieving lower inaccessibility rates than the human baseline (0.125). However, none of these techniques improved LLM-generated accessibility beyond \textit{Naive Code Generation}; instead, all resulted in slightly higher inaccessibility rates compared to the naive approach.

For GPT-4o, Zero-Shot prompting produced an inaccessibility rate of 0.112, which was lower than the human-generated rate (0.125) but still higher than \textit{Naive Code Generation} (0.105). Few-Shot and Self-Criticism exhibited similar trends, with inaccessibility rates of 0.118 and 0.112, respectively, showing no significant improvements over Zero-Shot.

For Qwen2.5-Coder, code generated with Zero-Shot (0.121) and Self-Criticism (0.122) were marginally more accessible than human-written code but still less accessible than \textit{Naive Code Generation} (0.113). While Few-Shot (0.111) performed slightly better than \textit{Naive Code Generation}, the improvement was marginal. This suggests that while prompting techniques slightly improved accessibility over human-written code, they did not enhance performance beyond Naive Generation.

We further analyzed the types of violations introduced by different prompting techniques. The detail analysis of Zero-Shot, Few-Shot, and Self-Criticism on accessibility violations are reported in the companion website~\cite{replicationpackage} for brevity.


\begin{tcolorbox}[colback=black!5!white,colframe=gray!75!black,fonttitle=\bfseries,title=Finding 2]
\textbf{Advanced prompting techniques consistently generate code with lower accessibility issues than human-written code, yet they fail to consistently surpass Naive Code Generation, indicating inherent limitations in addressing accessibility through prompting alone.}
\end{tcolorbox}

\subsubsection*{\underline{Discussion and Implications}}
Although structured prompting can effectively address certain basic accessibility issues, such as color contrast and alternative text descriptions, the inadvertent introduction of new ARIA-related violations reveals inherent limitations in explicit prompting. These results suggest that, despite their promise, prompting techniques must be cautiously designed to avoid overly rigid instructions that ignore context-specific constraints. Future work should consider developing more nuanced prompting approaches that dynamically adapt instructions based on contextual information to minimize unintended accessibility violations.

\subsection{RQ3: Can LLMs generate more accessible code by incorporating accessibility evaluation results into the code generation process?}
\label{subsec:rq3}
As discussed earlier in Section~\ref{subsec:feeda11y}, to address the limitations of prompting techniques in generating accessible code, we developed \textit{FeedA11y}. 
We now present our experimental evaluation of FeedA11y.

\subsubsection{Evaluation with AChecker.}
As shown in Table~\ref{tab:achecker_results_combined}, \textit{FeedA11y} achieved the lowest inaccessibility rate (0.300 with Qwen2.5-Coder), outperforming both human-written code (0.425) and Naive Code Generation (0.348). It surpassed humans and Naive Code Generation in 7 out of 10 projects.

Compared to human-written code and all other prompting methods, \textit{FeedA11y} with GPT-4o delivered substantial accessibility gains, notably reducing Contrast Issues by 53\% and improving key areas like \textit{Alternative Text Issues}, and \textit{Form Labeling Issues}. It also enhanced semantic structure adherence, including reductions in \textit{ARIA Tabbing Issues}. However, \textit{FeedA11y} still struggled with \textit{ARIA Landmark Uniqueness Issues}, suggesting room for refinement in handling complex ARIA patterns and also highlighting the need for additional targeted prompting or training strategies. Similar trends were observed with Qwen2.5-Coder, where \textit{FeedA11y} effectively reduced common issues and outperformed both humans and other prompting strategies, especially on basic accessibility principles and ARIA tabbing.

\subsubsection{Evaluation with QualWeb.}

As shown in Table~\ref{tab:qualweb_results_combined}, when GPT-4o was used for code generation, \textit{FeedA11y} achieved an inaccessibility rate of 0.112—lower than human-written code (0.1246) but slightly higher than Naive Code Generation (0.105). With Qwen2.5-Coder, however, \textit{FeedA11y} consistently outperformed all other techniques and human baselines, achieving the lowest inaccessibility rate.
While \textit{FeedA11y} with GPT-4o reduced issues like \textit{Contrast Issues} and \textit{Alternative Text}, it also introduced notable increases in \textit{Link Title Attribute} and \textit{List Link Groups violations}. Similar trends appeared with Qwen, where \textit{FeedA11y} improved on \textit{Contrast Issues} and \textit{Naive Code Generation} but raised other issues tied to semantic structure and interaction. Overall, while \textit{FeedA11y} demonstrated balanced improvements over humans and prompting techniques, challenges remain in mitigating certain semantic accessibility violations such as \textit{LinkTitleAttr} and \textit{ListLinkGroups}.

\begin{tcolorbox}[colback=black!5!white,colframe=gray!75!black,fonttitle=\bfseries,title=Finding 3]
\textbf{FeedA11y consistently outperforms human-written code and all prompting techniques in accessibility, especially when leveraging Qwen2.5-Coder.}
\end{tcolorbox}

\subsubsection*{\underline{Discussion and Implications}} 
These results underscore the effectiveness of leveraging evaluation-driven feedback within the code generation process to systematically reduce common accessibility violations. However, the persistent challenges with nuanced ARIA constraints, as well as accessibility issues related to semantic correctness (e.g., improper use of ARIA roles or missing landmark elements) and interactive behavior (e.g., inaccessible keyboard navigation or focus management), imply that evaluation-driven approaches alone are insufficient for managing all aspects of accessibility, particularly those involving complex semantics. Future enhancements could combine feedback-driven methods with specialized prompting techniques focusing on ARIA guidelines to improve accessibility outcomes.

\subsection{RQ4: To what extent does data leakage impact the effectiveness of \textit{FeedA11y} in generating accessible code?}
\label{subsec:data_leakage}

One potential threat to our experiment is the possibility that the LLMs used may have prior knowledge of the dataset, leading to unintended bias in the results. To mitigate this risk, we conduct a data leakage experiment for the LLMs used in our study. Specifically, we select two web projects~\cite{githubGitHubAlphaonelabseducationwebsite, githubGitHubNuranferhanecommercesite} that were created after the knowledge cutoff dates of Qwen2.5-Coder (September 2024) and GPT-4o (October 2023). These projects contained more than 30 UI-related files available for regeneration using \textit{FeedA11y}. 
Results show that both LLMs maintained lower inaccessibility rates (0.147 for GPT-4o and 0.155 for Qwen) than human-written code (0.225), consistent with the earlier findings. This confirms that LLM-generated accessibility improvements stem from model capabilities rather than memorization, strengthening the external validity of our study.

\begin{tcolorbox}[colback=black!5!white,colframe=gray!75!black,fonttitle=\bfseries,title=Finding 4]
\textbf{FeedA11y's performance in generating accessible code, surpassing human-written code, is not attributable to data leakage.}
\end{tcolorbox}

\subsection{RQ5: What design choices in \textit{FeedA11y} have the most significant impact on the accessibility of generated code?}
\label{subsec:ablation_study}

\textit{FeedA11y} can be configured in multiple ways depending on the choice of LLMs for code generation and optimization, as well as the accessibility guidelines used. To identify the most effective setup, we select the configuration with the lowest average inaccessibility rate compared to human-written code. This optimal setup uses Qwen2.5-Coder for generation, GPT-4o for optimization based on QualWeb rules, and AChecker for final evaluation. Using this configuration, we perform an ablation study to examine the impact of five key input features from Figure~\ref{fig:prompts_condensed} on code quality, which are \textit{Accessibility Instructions}, \textit{Guideline Descriptions}, \textit{Code Examples}, \textit{Testing Rules}, and \textit{Style Properties}. We exclude \textit{Multiple Rounds} and \textit{Multiple Agents} to focus on input features, while retaining \textit{Code Summaries} in all variants due to its central role in our approach.

Our ablation study shows that excluding Style Properties and  Accessibility Instructions led to the largest declines in accessibility, increasing the average inaccessibility rate by \textit{15.01\%} and \textit{14.83\%}, respectively. Removing Code Examples and Testing Rules also had notable negative effects, raising inaccessibility rates by \textit{2.23\%} and \textit{9.02\%}. In contrast, Guideline Descriptions had minimal impact, with only a \textit{0.99\%} increase in violations, indicating its relatively smaller role in accessibility improvements.

\begin{tcolorbox}[colback=black!5!white,colframe=gray!75!black,fonttitle=\bfseries,title=Finding 5]
\textbf{Style Properties and Accessibility Instructions are the two most significant components of the prompt.}
\end{tcolorbox}
\section{Threats to Validity}
\label{sec:threats}
While we employed rigorous methodological approaches in our study, we acknowledge certain considerations regarding validity.

\textbf{Construct Validity:}
Our use of inaccessibility rate as the primary metric poses certain limitations. Although it enables practical comparisons by measuring the proportion of elements with accessibility violations, it treats all violations equally, regardless of their severity or user impact. To address this, we provided a detailed breakdown of violation types and associated WCAG principles. Additionally, we employed both AChecker and QualWeb to mitigate tool-specific biases. However, this metric does not fully reflect the real user experience, which future work could address through studies involving users with disabilities.

\textbf{Internal Validity:}
To prevent information leakage between summarization and code generation, we employed separate LLM instances for each task. Additionally, we ran a data leakage experiment using projects released after the LLMs' knowledge cutoff dates to ensure no prior exposure. To mitigate subjectivity in manual webpage consistency verification, two authors independently evaluated outputs using predefined criteria and resolved discrepancies through a negotiated agreement process.

\textbf{External Validity:}
Our findings may not generalize beyond this study's LLMs and web projects. We mitigated this by selecting diverse projects from different domains with varying complexity and popularity.
\section{Conclusion}
\label{sec:conclusion}

In this paper, we systematically evaluated the accessibility of LLM-generated web code compared to human-written code across real-world projects. Our findings show that LLMs--particularly GPT-4o and Qwen2.5-Coder--often produce more accessible code than humans. While advanced prompting strategies (Zero-Shot, Few-Shot, Self-Criticism) improved accessibility, they also introduced new issues, especially with ARIA attribute uniqueness. To address these gaps, we introduced \textit{FeedA11y}, a ReAct-based technique that uses accessibility feedback to refine code iteratively, outperforming all other methods. Future work should focus on targeted strategies and training to help LLMs consistently generate fully accessible web content. 

We provide the source code and datasets that were used in our experiments on the companion website \cite{replicationpackage}.

\balance
\bibliographystyle{ACM-Reference-Format}
\bibliography{icse-accessibility}

\end{document}